\documentclass{article}%
\usepackage{amsmath}
\usepackage{amsfonts}
\usepackage{amssymb}
\usepackage{graphicx}%
\providecommand{\U}[1]{\protect\rule{.1in}{.1in}}

\begin{document}

\author{Giuseppe Castagnoli
\and Pieve Ligure, Italy, giuseppe.castagnoli@gmail.com}
\title{The quantum correlation between the selection of the problem and that of the
solution sheds light on the mechanism of the quantum speed up}
\maketitle

\begin{abstract}
In classical problem solving, there is of course correlation between the
selection of the problem on the part of Bob (the problem setter) and that of
the solution on the part of Alice (the problem solver). In quantum problem
solving, this correlation becomes quantum. This means that Alice contributes
to selecting 50\% of the information that specifies the problem. As the
solution is a function of the problem, this gives to Alice advanced knowledge
of 50\% of the information that specifies the solution. Both the quadratic and
exponential speed ups are explained by the fact that quantum algorithms start
from this advanced knowledge.

\end{abstract}

\section{Outline of the argument}

Quantum algorithms require fewer computation steps than their classical
counterparts. The reason for this quantum speed up is not well understood. For
example, recently Gross et al. $\left[  1\right]  $ asserted that the exact
reason for it has never been pinpointed. The key to the present explanation of
the speed up is the quantum correlation existing between the selection of the
problem on the part of Bob (the problem setter) and that of the solution on
the part of Alice (the problem solver). Because of it, all is like Alice
contributed to selecting 50\% of the information that specifies the problem.
Since the solution is a function of the problem, this gives to Alice the
advanced knowledge of 50\% of the information that specifies the solution. The
speed up is explained by the fact that Alice starts from this advanced knowledge.

The quantum correlation we are dealing with appears at the level of the whole
quantum experiment, from the measurement required to put the quantum system in
a known state, necessary to prepare the problem, to the measurement required
to extract the solution -- see also $\left[  2,3,4\right]  $.

We focus on Grover's $\left[  5\right]  $ quantum search algorithm. Bob
selects a value of $\mathbf{b}\in\left\{  0,1\right\}  ^{n}$, Alice should
find it by computing the Kronecker function $\delta\left(  \mathbf{b}%
,\mathbf{a}\right)  $ for various values of $\mathbf{a}\in\left\{
0,1\right\}  ^{n}$.\ We consider the simplest instance $n=2$. With a classical
algorithm, Alice should plan $3$ computations of $\delta\left(  \mathbf{b}%
,\mathbf{a}\right)  $ to be certain of finding the solution, with Grover's
algorithm, $1$ computation. There is a quantum speed up.

In the original Grover's algorithm, a register $A$, under the control of
Alice, contains the value of $\mathbf{a}$; the value of $\mathbf{b}$\ is
hard-wired inside the black box that, given in input a value of\ $\mathbf{a}$,
computes $\delta\left(  \mathbf{b},\mathbf{a}\right)  $. To highlight quantum
correlation, we add a register $B$, under the control of Bob, containing the
value of $\mathbf{b}$. We call $\hat{A}$ the content of register $A$, $\hat
{B}$ that of register $B$ -- $\hat{A}$ and $\hat{B}$ are commuting observables.

Initially, register $B$ is in a maximally mixed state. As usual, Alice
prepares register $A$ in a uniform, coherent superposition of all the possible
values of $\mathbf{a}$. The initial state of the two registers is thus:%

\begin{align}
\left\vert \psi\right\rangle  &  =\frac{1}{4}\left(  \operatorname{e}%
^{i\varphi_{0}}\left\vert 00\right\rangle _{B}+\operatorname{e}^{i\varphi_{1}%
}\left\vert 01\right\rangle _{B}+\operatorname{e}^{i\varphi_{2}}\left\vert
10\right\rangle _{B}+\operatorname{e}^{i\varphi_{3}}\left\vert 11\right\rangle
_{B}\right) \nonumber\\
&  \left(  \left\vert 00\right\rangle _{A}+\left\vert 01\right\rangle
_{A}+\left\vert 10\right\rangle _{A}+\left\vert 11\right\rangle _{A}\right)  ,
\label{in}%
\end{align}
where the $\varphi_{i}$ are independent random phases, each with uniform
distribution in $\left[  0,2\pi\right]  $. We use the random phase
representation of a mixed state, instead of the density operator, to keep the
usual state vector representation of the quantum algorithm. The density
operator is simply the average over all the $\varphi_{i}$ of the product of
the ket by the bra: $\left\langle \left\vert \psi\right\rangle \left\langle
\psi\right\vert \right\rangle _{\forall\varphi_{i}}$. The two bits von Neumann
entropy of the state of $B$ -- and of the overall quantum state (\ref{in}) --
corresponds to the complete indeterminacy of the value of $\mathbf{b}$.

In order to prepare register $B$ in the desired value of $\mathbf{b}$, say
$\mathbf{b}=00$, Bob measures $\hat{B}$ in state (\ref{in}), thus randomly
selecting a value of $\mathbf{b}$, say $\mathbf{b}=01$. This projects state
(\ref{in}) on:%
\begin{equation}
P_{\alpha}\left\vert \psi\right\rangle =\frac{1}{2}\left\vert 01\right\rangle
_{B}\left(  \left\vert 00\right\rangle _{A}+\left\vert 01\right\rangle
_{A}+\left\vert 10\right\rangle _{A}+\left\vert 11\right\rangle _{A}\right)  ,
\label{sul}%
\end{equation}
here and in the following we denote projection operators by the letter $P$.
The entropy of the quantum state goes to zero with the determination of the
value of $\mathbf{b}$. Then he applies to register $B$ a permutation of the
values of $\mathbf{b}$ -- a unitary transformation $U_{B}$ -- that changes the
randomly selected value into the desired one:%
\begin{equation}
U_{B}P_{\alpha}\left\vert \psi\right\rangle =\frac{1}{2}\left\vert
00\right\rangle _{B}\left(  \left\vert 00\right\rangle _{A}+\left\vert
01\right\rangle _{A}+\left\vert 10\right\rangle _{A}+\left\vert
11\right\rangle _{A}\right)  . \label{sulp}%
\end{equation}

The unitary part of the quantum algorithm, $U_{BA}$, sends state (\ref{sulp})
into%
\begin{equation}
U_{BA}U_{B}P_{\alpha}\left\vert \psi\right\rangle =\left\vert 00\right\rangle
_{B}\left\vert 00\right\rangle _{A}. \label{output}%
\end{equation}

Register $A$ contains the solution, namely the value of $\mathbf{b}$ chosen by
Bob. Alice acquires the solution by measuring $\hat{A}$. Of course there is a
one to one correlation between the value of $\mathbf{b}$ chosen by Bob and the
solution found by Alice. Up to the permutation introduced by $U_{B}$, this
corresponds to the quantum correlation between the outcome of measuring
$\hat{B}$ in (\ref{in}) and that of measuring $\hat{A}$ in (\ref{output}).
From the standpoint of quantum correlation, which concerns repetitions of the
same quantum experiment, $U_{B}$ should be considered fixed. The fact that Bob
chooses the permutation $U_{B}$ to obtain the desired value of $\mathbf{b}$
belongs to a different film.

With $U_{B}$ fixed, all is like Bob's measurement of $\hat{B}$ randomly
selected the value $\mathbf{b}=00$, which becomes in fact a fixed permutation
of the randomly selected value $\mathbf{b}=01$; in this special sense, we will
speak of the random selection also of the value of $\mathbf{b}$\ chosen by
Bob. Moreover, Bob's measurement can be performed indifferently at the
beginning or the end of the algorithm. We show the quantum development in the
case that this measurement is performed at the end:%
\begin{align}
\left\vert \psi\right\rangle  &  =\frac{1}{4}\left(  \operatorname{e}%
^{i\varphi_{0}}\left\vert 00\right\rangle _{B}+\operatorname{e}^{i\varphi_{1}%
}\left\vert 01\right\rangle _{B}+\operatorname{e}^{i\varphi_{2}}\left\vert
10\right\rangle _{B}+\operatorname{e}^{i\varphi_{3}}\left\vert 11\right\rangle
_{B}\right) \nonumber\\
&  \left(  \left\vert 00\right\rangle _{A}+\left\vert 01\right\rangle
_{A}+\left\vert 10\right\rangle _{A}+\left\vert 11\right\rangle _{A}\right)
\label{rel1}\\
U_{B}\left\vert \psi\right\rangle  &  =\left\vert \psi\right\rangle
\label{rel1bis}\\
U_{BA}U_{B}\left\vert \psi\right\rangle  &  =\frac{1}{2}\left(
\operatorname{e}^{i\varphi_{0}}\left\vert 00\right\rangle _{B}\left\vert
00\right\rangle _{A}+\operatorname{e}^{i\varphi_{1}}\left\vert 01\right\rangle
_{B}\left\vert 01\right\rangle _{A}+\operatorname{e}^{i\varphi_{2}}\left\vert
10\right\rangle _{B}\left\vert 10\right\rangle _{A}+\operatorname{e}%
^{i\varphi_{3}}\left\vert 11\right\rangle _{B}\left\vert 11\right\rangle
_{A}\right)  ,\label{rel3}\\
P_{\omega}U_{BA}U_{B}\left\vert \psi\right\rangle  &  =\left\vert
00\right\rangle _{B}\left\vert 00\right\rangle _{A}; \label{rel4}%
\end{align}
of course $U_{B}$\ changes the maximally mixed state of register $B$\ into
itself; for reasons that will become clear, we have assumed that the final
measurement of $\hat{A}$\ on the part of Alice still randomly projects on
$\mathbf{b}=00$. We can see why Bob's measurement can be deferred at the end:
the projection of (\ref{rel3}) on (\ref{rel4}), back evolved by $U_{B}^{\dag
}U_{BA}^{\dag}$, becomes the projection of (\ref{in}) on (\ref{sul}).

Thinking that all measurements are performed in the maximally entangled state
(\ref{rel3}) makes it more clear that the value of $\mathbf{b}$ is randomly
selected by either Bob's or Alice's measurement. Either measurement projects
state (\ref{rel3}) on the solution eigenstate (\ref{rel4}), where both
registers contain the randomly selected value of $\mathbf{b}$;
correspondingly, the 2 bit entropy of the quantum state goes to zero.

Unlike measurements, projections are not localized in time. They can be back
evolved by the inverse of the time forward unitary evolution. Thus, there is
no reason to ascribe the projection on the solution eigenstate, or the
corresponding reduction of entropy and random selection of the value of
$\mathbf{b}$, to one measurement rather than the other. Because of the
symmetry between the two measurements, we ascribe the determination of 50\% of
the bits of $\mathbf{b}$ to the measurement performed by Alice, the other 50\%
to that performed by Bob.

Halving the projection on the solution can be done in many ways. In section
2.3 we will symmetrize for all the possible ways, here we exemplify one way.
We break down $\hat{A}$ into content of first qubit $\hat{A}_{0}$ and content
of second qubit $\hat{A}_{1}$; we call $a_{0}$ ($a_{1}$) the eigenvalue
obtained by measuring $\hat{A}_{0}$ ($\hat{A}_{1}$). We define in a similar
way $\hat{B}_{0}$, $\hat{B}_{1}$, $b_{0}$, and $b_{1}$. We ascribe to Alice
the measurement of $\hat{A}_{0}$, which selects $a_{0}=b_{0}=0$, to Bob the
measurement of $\hat{B}_{1}$, which selects $a_{1}=b_{1}=0$. Together, the two
corresponding projections project on the solution; individually, they halve
the projection on the solution.

Summing up, half of the bits of $\mathbf{b}$ are randomly selected by Bob, the
other half by Alice. We show that this means that Alice knows in advance 50\%
of the bits of $\mathbf{b}$. It suffices to note that states (\ref{rel1})
through (\ref{rel4}) are the original quantum algorithm -- namely states
(\ref{in}) through (\ref{output}) -- with the quantum state relativized to the
observer Alice in the sense of relational quantum mechanics $\left[  6\right]
$. By definition, initially Alice does not know the content of register $B$.
To her, register $B$ is in a maximally mixed state even if Bob has already
measured $\hat{B}$. The 2 bit entropy of this state -- and of the overall
quantum state (\ref{rel1}) -- represents Alice's ignorance of the value of
$\mathbf{b}$. When Alice measures $\hat{A}$ at the end of the algorithm, the
quantum state (\ref{rel3}) is projected on the solution eigenstate
(\ref{rel4}). This projection is random to Alice, it is actually on the value
of $\mathbf{b}$\ chosen by Bob. The entropy of the quantum state goes to zero
and Alice acquires full knowledge of the value of $\mathbf{b}$. Thus, the
entropy of the relativized quantum state gauges Alice's ignorance of the value
of $\mathbf{b}$ throughout the execution of the algorithm.

As we have said before, when Alice measures $\hat{A}$ at the end of the
algorithm, half of the projection on the solution eigenstate is Alice's
contribution to the random selection of the value of $\mathbf{b}$. We back
evolve to the beginning of the quantum algorithm (to immediately after the
permutation $U_{B}$) this halved projection, for example the projection
associated with measuring $\hat{A}_{0}$ and obtaining $a_{0}=b_{0}=0$ -- we
should apply $U_{BA}^{\dag}$ to the projection. This projects the initial
state (\ref{rel1bis}) on%

\begin{equation}
\frac{1}{2\sqrt{2}}\left(  \operatorname{e}^{i\varphi_{0}}\left\vert
00\right\rangle _{B}+\operatorname{e}^{i\varphi_{1}}\left\vert 01\right\rangle
_{B}\right)  \left(  \left\vert 00\right\rangle _{A}+\left\vert
01\right\rangle _{A}+\left\vert 10\right\rangle _{A}+\left\vert
11\right\rangle _{A}\right)  , \label{am}%
\end{equation}
halving the entropy of the state of register $B$. This means that Alice,
before starting the algorithm and "after" this back evolved half projection,
knows that $b_{0}=0$, namely one of the two bits of the solution she will read
in the future in register $A$.

We are at the level of elementary logical operations, where knowing means
doing. Alice knows of the advanced information by acting like she knew it,
namely by using it to identify classically the missing bit (the value of
$b_{1}$) with a single computation of $\delta\left(  \mathbf{b},\mathbf{a}%
\right)  $. Correspondingly, as we showed in $\left[  3,4\right]  $, the
quantum algorithm is the superposition of all the possible ways of taking one
bit of information about the solution and, given the advanced knowledge of
this bit, classically identifying the missing bit with a single computation of
$\delta\left(  \mathbf{b},\mathbf{a}\right)  $. This explains the speed up
from 3 to 1 computation.

We note that the entangled state (\ref{rel3}) is the outcome of the unitary
part of any quantum algorithm that starts with a maximally mixed state of
register $B$\ and solves the data base search problem, with or without a
quantum speed up. In fact the quantum algorithm can do either without or with
the advanced knowledge. In the former case, it is isomorphic with a classical
algorithm that starts from the usual initial state and yields no speed up. In
the latter, it is isomorphic with a classical algorithm that starts from the
initial state "after" the back evolved half projection on the solution -- thus
with advanced knowledge of 50\% of the bits of the solution.

The above explanation of the speed up\ generalizes to $n>2$ and to the very
diverse quantum algorithms that yield an exponential speed-up. In all the
cases examined, the quantum algorithm requires the number of function
evaluations (computations of $\delta\left(  \mathbf{b},\mathbf{a}\right)
$\ in Grover's case) of a classical algorithm that knows in advance 50\% of
the information about the solution. Already in former work $\left[  3\right]
$, we called this \textit{the 50\% rule of the quantum algorithms}.

The 50\% rule has a practical interest, it allows to characterize the problems
solvable with a quantum speed up in an entirely computer science framework
with no physics involved -- an important simplification. It should also allow
to identify new quantum speed ups.

Section 2 highlights the mechanism of the speed up in the case of Grover's
algorithm. In section 3, we check that the 50\% rule holds for a class of
quantum algorithms that yield an exponential speed-up. In section 4, we
develop a new quantum speed up out of the 50\% rule. In section 5 we draw the conclusions.

\section{The mechanism of the quantum speed-up in Grover's algorithm}

We develop in detail the line of thinking provided in the former section,
first for $n=2$ then for $n>2$.

\subsection{ Extended representation of Grover's algorithm}

We relativize to Alice the quantum state of the original Grover's algorithm.
With $n=2$, registers $B$ and $A$ are two-qubits each. A one-qubit register
$V$ is meant to contain the result of the computation of $\delta\left(
\mathbf{b},\mathbf{a}\right)  $, modulo 2 added to its initial content for
logical reversibility. Let us assume that Bob chose $\mathbf{b}=00$, the
initial state of the three registers is anyhow:%

\begin{align}
\left\vert \Psi\right\rangle  &  =\frac{1}{4\sqrt{2}}\left(  \operatorname{e}%
^{i\varphi_{0}}\left\vert 00\right\rangle _{B}+\operatorname{e}^{i\varphi_{1}%
}\left\vert 01\right\rangle _{B}+\operatorname{e}^{i\varphi_{2}}\left\vert
10\right\rangle _{B}+\operatorname{e}^{i\varphi_{3}}\left\vert 11\right\rangle
_{B}\right) \label{inv}\\
&  \left(  \left\vert 00\right\rangle _{A}+\left\vert 01\right\rangle
_{A}+\left\vert 10\right\rangle _{A}+\left\vert 11\right\rangle _{A}\right)
\left(  \left\vert 0\right\rangle _{V}-\left\vert 1\right\rangle _{V}\right)
.\nonumber
\end{align}
The two bit entropy of the state of register $B$ represents Alice's initial
ignorance of Bob's choice.

The computation of $\delta\left(  \mathbf{b},\mathbf{a}\right)  $ is performed
in quantum parallelism on each term of the superposition. For example, the
input $\operatorname{e}^{i\varphi_{1}}\left\vert 01\right\rangle
_{B}\left\vert 01\right\rangle _{A}\left\vert 0\right\rangle _{V}$ means that
the input of the computation of $\delta\left(  \mathbf{b},\mathbf{a}\right)
$\ is $\mathbf{b}=01,$ $\mathbf{a}=01$ and that the initial content of
register $V$\ is $0$. The computation yields $\delta\left(  01,01\right)  =1$
that, modulo 2 added to the initial content of $V$, yields the output
$\operatorname{e}^{i\varphi_{1}}\left\vert 01\right\rangle _{B}\left\vert
01\right\rangle _{A}\left\vert 1\right\rangle _{V}$ ($B$\ and $A$ keep the
memory of the input for logical reversibility). Similarly, the input
$-\operatorname{e}^{i\varphi_{1}}\left\vert 01\right\rangle _{B}\left\vert
01\right\rangle _{A}\left\vert 1\right\rangle _{V}$ goes into the output
$-\operatorname{e}^{i\varphi_{1}}\left\vert 01\right\rangle _{B}\left\vert
01\right\rangle _{A}\left\vert 0\right\rangle _{V}$. More in general, the
input $\operatorname{e}^{i\varphi_{\mathbf{b}}}\left\vert \mathbf{b}%
\right\rangle _{B}\left\vert \mathbf{a}\right\rangle _{A}\left(  \left\vert
0\right\rangle _{V}-\left\vert 1\right\rangle _{V}\right)  $ goes into the
output $-\operatorname{e}^{i\varphi_{\mathbf{b}}}\left\vert \mathbf{b}%
\right\rangle _{B}\left\vert \mathbf{a}\right\rangle _{A}\left(  \left\vert
0\right\rangle _{V}-\left\vert 1\right\rangle _{V}\right)  $ if $\mathbf{a}%
=\mathbf{b}$, remains unaltered otherwise.\ In the overall, a single
computation of $\delta\left(  \mathbf{b},\mathbf{a}\right)  $ sends state
(\ref{inv}) into:%

\begin{equation}
U_{\delta}\left\vert \Psi\right\rangle =\frac{1}{4\sqrt{2}}\left[
\begin{array}
[c]{c}%
\operatorname{e}^{i\varphi_{0}}\left\vert 00\right\rangle _{B}\left(
-\left\vert 00\right\rangle _{A}+\left\vert 01\right\rangle _{A}+\left\vert
10\right\rangle _{A}+\left\vert 11\right\rangle _{A}\right)  +\\
\operatorname{e}^{i\varphi_{1}}\left\vert 01\right\rangle _{B}\left(
\left\vert 00\right\rangle _{A}-\left\vert 01\right\rangle _{A}+\left\vert
10\right\rangle _{A}+\left\vert 11\right\rangle _{A}\right)  +\\
\operatorname{e}^{i\varphi_{2}}\left\vert 10\right\rangle _{B}\left(
\left\vert 00\right\rangle _{A}+\left\vert 01\right\rangle _{A}-\left\vert
10\right\rangle _{A}+\left\vert 11\right\rangle _{A}\right)  +\\
\operatorname{e}^{i\varphi_{3}}\left\vert 11\right\rangle _{B}\left(
\left\vert 00\right\rangle _{A}+\left\vert 01\right\rangle _{A}+\left\vert
10\right\rangle _{A}-\left\vert 11\right\rangle _{A}\right)
\end{array}
\right]  \left(  \left\vert 0\right\rangle _{V}-\left\vert 1\right\rangle
_{V}\right)  , \label{int}%
\end{equation}
a maximally entangled state where four orthogonal states of register $B$, each
containing a single value of $\mathbf{b}$,\ are correlated with four
orthogonal states of register $A$. To transform entanglement into correlation
between measurement outcomes, we apply to register $A$ the unitary
transformation $U_{A}$ such that:%

\begin{align}
U_{A}U_{\delta}\left\vert \Psi\right\rangle  &  =\frac{1}{2\sqrt{2}}\left(
\operatorname{e}^{i\varphi_{0}}\left\vert 00\right\rangle _{B}\left\vert
00\right\rangle _{A}+\operatorname{e}^{i\varphi_{1}}\left\vert 01\right\rangle
_{B}\left\vert 01\right\rangle _{A}+\operatorname{e}^{i\varphi_{2}}\left\vert
10\right\rangle _{B}\left\vert 10\right\rangle _{A}+\operatorname{e}%
^{i\varphi_{3}}\left\vert 11\right\rangle _{B}\left\vert 11\right\rangle
_{A}\right) \label{outv}\\
&  \left(  \left\vert 0\right\rangle _{V}-\left\vert 1\right\rangle
_{V}\right)  .\nonumber
\end{align}
We incidentally note that eliminating register $V$, like we did in section 1,
does not alter the unitary character of the transformations.\ Measuring
$\hat{A}$ in state (\ref{outv}), projects it on the solution eigenstate:%
\begin{equation}
\frac{1}{\sqrt{2}}\left\vert 00\right\rangle _{B}\left\vert 00\right\rangle
_{A}\left(  \left\vert 0\right\rangle _{V}-\left\vert 1\right\rangle
_{V}\right)  , \label{eigenv}%
\end{equation}
yielding the eigenvalue $\mathbf{a}=00$, namely the solution of the problem.
Alice acquires full knowledge of the value of $\mathbf{b}$ chosen by Bob and
the entropy of the quantum state becomes zero. This entropy gauges Alice's
knowledge of the value of $\mathbf{b}$ throughout the execution of the algorithm.

\subsection{Back evolving 50\% of the projection on the solution}

We show the consequence of ascribing 50\% of the determination of the value of
$\mathbf{b}$ to a partial measurement performed by Alice\footnote{We should
keep in mind that Alice's measurement contributes to the random selection of a
value of $\mathbf{b}$, then transformed into the value chosen by Bob\ by the
unitary transformation $U_{B}$. Since this latter should be considered fixed
from the standpoint of quantum correlation, we can say that Alice contributes
to Bob's choice.}. We adopt the example of section 1; we assume that the value
of $\mathbf{b}$\ chosen by Bob is $\mathbf{b}=00$ and that the partial
measurement is that of $\hat{A}_{0}$.\ This selects the eigenvalue $a_{0}=0$,
projecting (\ref{outv}) on:%

\begin{equation}
\frac{1}{2}\left(  \operatorname{e}^{i\varphi_{0}}\left\vert 00\right\rangle
_{B}\left\vert 00\right\rangle _{A}+\operatorname{e}^{i\varphi_{1}}\left\vert
01\right\rangle _{B}\left\vert 01\right\rangle _{A}\right)  \left(  \left\vert
0\right\rangle _{V}-\left\vert 1\right\rangle _{V}\right)  . \label{poutv}%
\end{equation}
We back evolve this projection to the beginning of the quantum algorithm, by
applying to state (\ref{poutv}) the inverse of the unitary part of the
algorithm, namely $U_{\delta}^{\dag}U_{A}^{\dag}$. This projects the initial
state of the algorithm, (\ref{inv}), on:%

\begin{equation}
\frac{1}{4}\left(  \operatorname{e}^{i\varphi_{0}}\left\vert 00\right\rangle
_{B}+\operatorname{e}^{i\varphi_{1}}\left\vert 01\right\rangle _{B}\right)
\left(  \left\vert 00\right\rangle _{A}+\left\vert 01\right\rangle
_{A}+\left\vert 10\right\rangle _{A}+\left\vert 11\right\rangle _{A}\right)
\left(  \left\vert 0\right\rangle _{V}-\left\vert 1\right\rangle _{V}\right)
. \label{amv}%
\end{equation}

That the state of register $B$\ should have the form it has in (\ref{amv}),
can be seen more directly as follows. We note that the unitary part of the
quantum algorithm is the identity on the reduced density operator of register
$B$ that, in the random phase representation, is $\rho_{B}=\frac{1}{2}\left(
\operatorname{e}^{i\varphi_{0}}\left\vert 00\right\rangle _{B}%
+\operatorname{e}^{i\varphi_{1}}\left\vert 01\right\rangle _{B}%
+\operatorname{e}^{i\varphi_{2}}\left\vert 10\right\rangle _{B}%
+\operatorname{e}^{i\varphi_{3}}\left\vert 11\right\rangle _{B}\right)  $ in
both (\ref{inv}) and (\ref{outv}). By measuring $\hat{A}_{0}$\ in state
(\ref{outv}), Alice projects $\rho_{B}$ on $\frac{1}{\sqrt{2}}\left(
\operatorname{e}^{i\varphi_{0}}\left\vert 00\right\rangle _{B}%
+\operatorname{e}^{i\varphi_{1}}\left\vert 01\right\rangle _{B}\right)  $.
This projection goes back unaltered to the beginning of the algorithm.

State (\ref{amv}) says that, "after" back evolved projection, Alice knows in
advance that the value of $\mathbf{b}$ is either $\mathbf{b}=00$ or
$\mathbf{b}=01$, namely that $b_{0}=0$. Correspondingly, the entropy
representing Alice's initial ignorance of the solution has decreased from two
to one bit. How Alice utilizes this gain in information to achieve a speed up
is explained in the next section.

\subsection{Utilizing the back evolved half projections}

By measuring $\hat{A}$\ Alice projects state (\ref{outv}) on the solution
eigenstate. Let us go exhaustively through all the possible ways of halving
this projection on the solution. Until now we have considered the binary
observable $\hat{A}_{0}$, whose measurement tells whether $\mathbf{b}%
\in\left\{  00,01\right\}  $ or $\mathbf{b}\in\left\{  10,11\right\}  $, and
$\hat{A}_{1}$, whose measurement tells whether $\mathbf{b}\in\left\{
00,10\right\}  $ or $\mathbf{b}\in\left\{  01,11\right\}  $. There is a third
binary observable, say $\hat{A}_{+}$, whose measurement tells whether
$\mathbf{b}\in\left\{  00,11\right\}  $ or $\mathbf{b}\in\left\{
01,10\right\}  $. Measuring any pair of these three observables projects the
output state (\ref{outv})\ on the solution. Measuring any single observable
halves the projection on the solution.

In the overall, there are $6$ halved projections, on: $\left\{  00,01\right\}
$, $\left\{  10,11\right\}  $, ..., and $\left\{  01,10\right\}  $ -- all the
ways of taking a pair of elements out of four. Each halved projection
(actually, on an incoherent superposition of two values of $\mathbf{b}$) goes
back unaltered to the beginning of the quantum algorithm, where it halves the
entropy of Alice's state of knowledge of the value of $\mathbf{b}$,
originating 8 classical computation histories, as follows.

Let us start with the projection on $\mathbf{b}\in\left\{  00,01\right\}  $.
In other words, Alice knows in advance that $\mathbf{b}\in\left\{
00,01\right\}  $. To identify the missing bit, she should compute
$\delta\left(  \mathbf{b},\mathbf{a}\right)  $ for either $\mathbf{a}=00$ or
$\mathbf{a}=01$. We assume that she does it for $\mathbf{a}=00$ -- we are
pinpointing one of the possible combinations. If the outcome of the
computation is $\delta=1$, this means that $\mathbf{b}=00$. This originates
two classical computation histories (represented as sequences of sharp quantum
states), depending on the initial state of register $V$. History \# 1: initial
state $\operatorname{e}^{i\varphi_{0}}\left\vert 00\right\rangle
_{B}\left\vert 00\right\rangle _{A}\left\vert 0\right\rangle _{V}$, state
after the computation $\operatorname{e}^{i\varphi_{0}}\left\vert
00\right\rangle _{B}\left\vert 00\right\rangle _{A}\left\vert 1\right\rangle
_{V}$. History \# 2: initial state $\operatorname{e}^{i\varphi_{0}}\left\vert
00\right\rangle _{B}\left\vert 00\right\rangle _{A}\left\vert 1\right\rangle
_{V}$, state after the computation $\operatorname{e}^{i\varphi_{0}}\left\vert
00\right\rangle _{B}\left\vert 00\right\rangle _{A}\left\vert 0\right\rangle
_{V}$.\ If the outcome of the computation is $\delta=0$, this means that
$\mathbf{b}=01$. This originates other two histories. History \# 3: initial
state $\operatorname{e}^{i\varphi_{1}}\left\vert 01\right\rangle
_{B}\left\vert 00\right\rangle _{A}\left\vert 0\right\rangle _{V}$, state
after the computation $\operatorname{e}^{i\varphi_{1}}\left\vert
01\right\rangle _{B}\left\vert 00\right\rangle _{A}\left\vert 0\right\rangle
_{V}$. History \# 4: initial state $\operatorname{e}^{i\varphi_{1}}\left\vert
01\right\rangle _{B}\left\vert 00\right\rangle _{A}\left\vert 1\right\rangle
_{V}$, state after the computation $\operatorname{e}^{i\varphi_{1}}\left\vert
01\right\rangle _{B}\left\vert 00\right\rangle _{A}\left\vert 1\right\rangle
_{V}$.\ If she computes $\delta\left(  \mathbf{b},\mathbf{a}\right)  $ for
$\mathbf{a}=01$ instead, this originates other 4 histories. Etc.

If we sum together all the different histories (some histories are originated
more than once), each with a suitable phase, and normalize, we obtain the
function evaluation stage of the quantum algorithm, namely the transformation
of state (\ref{inv})\ into (\ref{int}).

This answers the question of how Alice knows of the \textit{advanced
information} -- the information conveyed back by the back evolved half
projections on the solution. We are at the level of elementary logical
operations, where "knowing" means "doing". Alice knows of the advanced
information by acting like she knew it, namely by computing on the basis of it
the missing information. It should be noted that Alice could also ignore (do
without) the advanced information, which simply means a quantum algorithm with
no speed up, isomorphic with a classical algorithm that starts from complete
ignorance of the value of $\mathbf{b}$. An algorithm that yields a speed up is
isomorphic with a classical algorithm that starts from the back evolved half
projections on the solution.

The 50\% rule only says that the quantum algorithm can be broken down into a
superposition of classical computation histories that start from the advanced
information, the history phases and the rotation of the basis of register $A$
(i. e. $U_{A}$) after the computation of $\delta\left(  \mathbf{b}%
,\mathbf{a}\right)  $ are what is needed for reconstructing the quantum
algorithm. However, in Ref. $\left[  3,4\right]  $, we have shown that the
quantum algorithm can be synthesized out of the advanced information classical
algorithm (out of the classical computation histories in quantum notation)
through an optimization procedure. We should choose history phases and
rotation of the basis of $A$ in such a way that they maximize: (i)
entanglement between registers $A$ and $B$ after the computation of
$\delta\left(  \mathbf{b},\mathbf{a}\right)  $ or, in equivalent terms, (ii)
the information about the solution readable in $A$\ at the end of the algorithm.

\subsection{Quantum search for $n>2$}

Registers $B$\ and $A$\ are $n$-qubit each. Register $V$\ is one-qubit. Given
the advanced knowledge of $n/2$ of the bits of the value of $\mathbf{b}$
selected by Bob, in order to compute the missing $n/2$ bits, Alice should
compute $\delta\left(  \mathbf{b},\mathbf{a}\right)  $ for all the values of
$\mathbf{a}$\ in quantum superposition\ and apply to register $A$\ the
appropriate unitary transformation $U_{A}$\ an\ $\operatorname{O}\left(
2^{n/2}\right)  $ times; each time $U_{A}$ maximizes the entanglement between
registers $B$ and $A$. Eventually we obtain (approximately):%

\begin{equation}
\frac{1}{2^{\left(  n+1\right)  /2}}\left(  \sum_{\mathbf{c}=0}^{2^{n}%
-1}\operatorname{e}^{i\varphi_{\mathbf{c}}}\left\vert \mathbf{c}\right\rangle
_{B}\left\vert \mathbf{c}\right\rangle _{A}\right)  \left(  \left\vert
0\right\rangle _{V}-\left\vert 1\right\rangle _{V}\right)  . \label{outgen}%
\end{equation}
Measuring either $\hat{A}$ or $\hat{B}$, or both, projects (\ref{outgen}) on
the solution eigenstate. According to the rationale of the previous sections,
we should halve the final projection on the solution in all possible ways; for
example, by measuring $\hat{A}_{0},~...,~\hat{A}_{\frac{n}{2}-1}$. Let
$\mathcal{I}$\ be the information acquired by reading the solution at the end
of the algorithm. Evidently, the considerations of the previous sections apply
also here: back evolving a half projection to the beginning of the quantum
algorithm, makes available at the input of the computation the corresponding
50\% of $\mathcal{I}$.

The fact that, for large $n$, the optimal number of times is $\frac{\pi}%
{4}2^{n/2}$, not $2^{n/2}$, does not imply that Grover's algorithm outperforms
the 50\% rule. In fact this optimal number is associated with a non-zero
probability -- $\operatorname{O}\left(  1/2^{n}\right)  $ -- that the
algorithm delivers a wrong solution. One should look for the possible cases
where Grover's algorithm yields the solution with certainty, like in the case
$n=2$.

\section{Checking the 50\% rule on other quantum algorithms}

Until now we have discussed the 50\% rule on Grover's algorithm. It is
therefore important to check that the rule holds for the very diverse quantum
algorithms that yield an exponential speed-up. In many of these algorithms,
there is a set of functions $f_{\mathbf{b}}:\left\{  0,1\right\}
^{n}\rightarrow\left\{  0,1\right\}  ^{m}$ known to both Alice and Bob. Bob
selects a value of $\mathbf{b}$ and Alice should find a character of the
function $f_{\mathbf{b}}$ by computing $f_{\mathbf{b}}\left(  \mathbf{a}%
\right)  $\ for various values of $\mathbf{a}$. Since the problems addressed
by such algorithms are structured, identifying the advanced information and
sharing out the projection on the solution between Alice and Bob requires some
care. With respect to the similar section of Ref. $\left[  4\right]  $, the
present one provides various clarifications.

\subsection{Deutsch\&Jozsa's algorithm}

In Deutsch\&Jozsa's $\left[  7\right]  $ algorithm, the set of functions known
to both Bob and Alice is all the constant and "balanced" functions (with an
even number of zeroes and ones) $f_{\mathbf{b}}:\left\{  0,1\right\}
^{n}\rightarrow\left\{  0,1\right\}  $. Table (\ref{dj}) gives this set for
$n=2$. The string $\mathbf{b}\equiv b_{0},b_{1},...,b_{2^{n}-1}$ is both the
suffix and the table of the function -- the sequence of function values for
increasing values of the argument.
\begin{equation}%
\begin{tabular}
[c]{|c|c|c|c|c|c|c|c|c|}\hline
$\mathbf{a}$ & $\,f_{0000}\left(  \mathbf{a}\right)  $ & $f_{1111}\left(
\mathbf{a}\right)  $ & $f_{0011}\left(  \mathbf{a}\right)  $ & $f_{1100}%
\left(  \mathbf{a}\right)  $ & $f_{0101}\left(  \mathbf{a}\right)  $ &
$f_{1010}\left(  \mathbf{a}\right)  $ & $f_{0110}\left(  \mathbf{a}\right)  $
& $f_{1001}\left(  \mathbf{a}\right)  $\\\hline
00 & 0 & 1 & 0 & 1 & 0 & 1 & 0 & 1\\\hline
01 & 0 & 1 & 0 & 1 & 1 & 0 & 1 & 0\\\hline
10 & 0 & 1 & 1 & 0 & 0 & 1 & 1 & 0\\\hline
11 & 0 & 1 & 1 & 0 & 1 & 0 & 0 & 1\\\hline
\end{tabular}
\ \label{dj}%
\end{equation}

Alice should find whether the function selected by Bob is balanced or
constant, by computing $f_{\mathbf{b}}\left(  \mathbf{a}\right)  =f\left(
\mathbf{b},\mathbf{a}\right)  $. In the classical case this requires, in the
worst case, a number of computations of $f\left(  \mathbf{b},\mathbf{a}%
\right)  $ exponential in $n$; in the quantum case one computation.

The initial state of the algorithm relativized to Alice is:%

\begin{align}
&  \frac{1}{2\sqrt{2}}\left(  \rho_{0}\operatorname{e}^{i\varphi_{0}%
}\left\vert 0000\right\rangle _{B}+\rho_{1}\operatorname{e}^{i\varphi_{1}%
}\left\vert 1111\right\rangle _{B}+\rho_{2}\operatorname{e}^{i\varphi_{2}%
}\left\vert 0011\right\rangle _{B}+\rho_{3}\operatorname{e}^{i\varphi_{3}%
}\left\vert 1100\right\rangle _{B}+...\right) \label{indj}\\
&  \left(  \left\vert 00\right\rangle _{A}+\left\vert 01\right\rangle
_{A}+\left\vert 10\right\rangle _{A}+\left\vert 11\right\rangle _{A}\right)
\left(  \left\vert 0\right\rangle _{V}-\left\vert 1\right\rangle _{V}\right)
.\nonumber
\end{align}
The coefficients $\rho_{i}>0$, such that $\sum\rho_{i}^{2}=1$, account for a
non flat probability distribution of Bob's selection. Things will be simpler
if we assume that the $\rho_{i}$ are the same for \textit{dual} values of
$\mathbf{b}$, like $0000$ and $1111$. Modulo $2$\ adding the result of the
computation of $f\left(  \mathbf{b},\mathbf{a}\right)  $ to the content of $V$
and performing the Hadamard transform on register $A$ yields the entangled state:%

\begin{align}
&  \frac{1}{\sqrt{2}}\left[  \left(  \rho_{0}\operatorname{e}^{i\varphi_{0}%
}\left\vert 0000\right\rangle _{B}-\rho_{1}\operatorname{e}^{i\varphi_{1}%
}\left\vert 1111\right\rangle _{B}\right)  \left\vert 00\right\rangle
_{A}+\left(  \rho_{2}\operatorname{e}^{i\varphi_{2}}\left\vert
0011\right\rangle _{B}-\rho_{3}\operatorname{e}^{i\varphi_{3}}\left\vert
1100\right\rangle _{B}\right)  \left\vert 10\right\rangle _{A}+....\right]
\label{hdj}\\
&  \left(  \left\vert 0\right\rangle _{V}-\left\vert 1\right\rangle
_{V}\right)  ,\nonumber
\end{align}
Measuring $\hat{B}$ and $\hat{A}$\ in (\ref{hdj}) yields Bob's selection of a
value of $\mathbf{b}$ and the solution found by Alice: all zeroes if the
function is constant, not so if it is balanced.

We check that the quantum algorithm requires the number of function
evaluations of a classical algorithm that knows in advance 50\% of
$\mathcal{I}$ -- we call $\mathcal{I}$ the information acquired by reading the
solution at the end of the algorithm. Since the solution is a function of
$\mathbf{b}$, we can define the advanced information as any 50\% of the
information about the solution contained in $\mathbf{b}$, namely in the table
of $f_{\mathbf{b}}\left(  \mathbf{a}\right)  $. If $f_{\mathbf{b}}\left(
\mathbf{a}\right)  $ is constant, for reasons of symmetry, the advanced
information is any 50\% of the table of the function -- see table (\ref{dj}).
If the function is balanced, still for reasons of symmetry, it is any 50\% of
the table that does not contain different values of the function -- for each
balanced function there are two such half tables. In fact, the half tables
that contain different values of the function already tell that the function
is balanced and thus contain 100\% of $\mathcal{I}$. For the \textit{good}
half tables, that do not contain different values of the function, the
solution (whether the function is constant or balanced) is always identified
by computing $f_{\mathbf{b}}\left(  \mathbf{a}\right)  $ for only one value of
$\mathbf{a}$\ (any one) outside the half table. Thus, both the quantum
algorithm and the advanced information classical algorithm require just one
function evaluation.

We should note that the present definition of advanced information:

\begin{enumerate}
\item Could be applied as well to Grover's algorithm, where it becomes: any
50\% of the table of $\delta\left(  \mathbf{b},\mathbf{a}\right)  $ (for a
given value of $\mathbf{b}$) that does not contain the value $\delta\left(
\mathbf{b},\mathbf{a}\right)  =1$. All the results of section 2 would remain unaltered.

\item Identifies a back evolved half projection on the solution. In fact -- up
to the sign of the random phase factors that is irrelevant -- the reduced
density operator of register $B$ in the random phase representation is:%
\begin{equation}
\rho_{B}=\rho_{0}\operatorname{e}^{i\varphi_{0}}\left\vert 0000\right\rangle
_{B}+\rho_{1}\operatorname{e}^{i\varphi_{1}}\left\vert 1111\right\rangle
_{B}+\rho_{2}\operatorname{e}^{i\varphi_{2}}\left\vert 0011\right\rangle
_{B}+\rho_{3}\operatorname{e}^{i\varphi_{3}}\left\vert 1100\right\rangle
_{B}+... \label{addj}%
\end{equation}
throughout the unitary part of the quantum algorithm. We assume that the
advanced information (a good half table) is $f\left(  \mathbf{b},00\right)
=0$ and $f\left(  \mathbf{b},01\right)  =0$. This means that the function
selected by Bob is either $\,f_{0000}\left(  \mathbf{a}\right)  $ or
$f_{0011}\left(  \mathbf{a}\right)  $ -- see table (\ref{dj}). This
corresponds to projecting $\rho_{B}$\ on $\rho_{B}^{\prime}=\left(  \rho
_{0}\operatorname{e}^{i\varphi_{0}}\left\vert 0000\right\rangle _{B}+\rho
_{2}\operatorname{e}^{i\varphi_{2}}\left\vert 0011\right\rangle _{B}\right)
$, up to normalization; this outcome goes back unaltered to the beginning of
the quantum algorithm, where it becomes Alice's advanced knowledge of the
solution. We should note that Alice, by measuring $\hat{A}_{1}$ in state
(\ref{hdj}) and finding $a_{1}=0$, projects $\rho_{B}$ not on $\rho
_{B}^{\prime}$ but on:%
\begin{equation}
\rho_{B}=\rho_{0}\operatorname{e}^{i\varphi_{0}}\left\vert 0000\right\rangle
_{B}+\rho_{1}\operatorname{e}^{i\varphi_{1}}\left\vert 1111\right\rangle
_{B}+\rho_{2}\operatorname{e}^{i\varphi_{2}}\left\vert 0011\right\rangle
_{B}+\rho_{3}\operatorname{e}^{i\varphi_{3}}\left\vert 1100\right\rangle _{B},
\label{dis}%
\end{equation}
up to normalization. To project (\ref{dis}) on $\rho_{B}^{\prime}$, Bob should
measure a single $\hat{B}_{i}$, e. g. $\hat{B}_{0}$, thus finding in present
assumptions $b_{0}=0$. This latter projection, although performed by Bob, can
be added to Alice's advanced knowledge of the solution. In fact it selects
between dual values of $\mathbf{b}$, which does not disclose to Alice any
information about the solution -- does not affect the entropy of the reduced
density operator of register $A$ in state (\ref{hdj}).
\end{enumerate}

This time, sharing out the projection on the solution between Alice and Bob
would be more complex, because of the asymmetry between the two actions.
However, we can bypass this difficulty. It suffices to note that, with all the
$\rho_{i}>0$, state (\ref{hdj}) is certainly entangled. Thus, in present
criteria, there is anyhow a non zero contribution to the determination of the
value of $\mathbf{b}$ on the part of both Alice and Bob. This is enough to see
that the advanced information available to Alice cannot exceed 50\% of
$\mathcal{I}$. In fact, increasing it over 50\% would mean increasing any good
half table by one row, which would project the output state (\ref{hdj}) on the
solution, leaving to Bob nothing to project.

Summing up, we have ascertained that Alice's advanced information is back
evolved projection and that it is (and cannot exceed) 50\% of $\mathcal{I}$.\ 

Now we go to the history superposition picture. Let us assume that the
advanced information is $f\left(  \mathbf{b},00\right)  =0$ and $f\left(
\mathbf{b},01\right)  =0$. Alice can find the value of $\mathbf{b}$ (thus the
character of the function), by performing function evaluation for either
$\mathbf{a}=10$ or $\mathbf{a}=11$. We assume that she does it for
$\mathbf{a}=10$. If the result of the computation is $0$, \ this means that
$\mathbf{b}=0000$. This originates two classical computation histories in
quantum notation: \# 1: initial state $\rho_{0}\operatorname{e}^{i\varphi_{0}%
}\left\vert 0000\right\rangle _{B}\left\vert 10\right\rangle _{A}\left\vert
0\right\rangle _{V}$, state after the computation $\rho_{0}\operatorname{e}%
^{i\varphi_{0}}\left\vert 0000\right\rangle _{B}\left\vert 10\right\rangle
_{A}\left\vert 0\right\rangle _{V}$; \# 2: initial state $\rho_{0}%
\operatorname{e}^{i\varphi_{0}}\left\vert 0000\right\rangle _{B}\left\vert
10\right\rangle _{A}\left\vert 1\right\rangle _{V}$, state after the
computation $\rho_{0}\operatorname{e}^{i\varphi_{0}}\left\vert
0000\right\rangle _{B}\left\vert 10\right\rangle _{A}\left\vert 1\right\rangle
_{V}$. If the result of the computation is $1$, this means that $\mathbf{b}%
=0011$. This originates two histories: \# 3: initial state $\rho
_{2}\operatorname{e}^{i\varphi_{2}}\left\vert 0011\right\rangle _{B}\left\vert
10\right\rangle _{A}\left\vert 0\right\rangle _{V}$, state after the
computation $\rho_{2}\operatorname{e}^{i\varphi_{2}}\left\vert
0011\right\rangle _{B}\left\vert 10\right\rangle _{A}\left\vert 1\right\rangle
_{V}$; \# 4: initial state $\rho_{2}\operatorname{e}^{i\varphi_{2}}\left\vert
0011\right\rangle _{B}\left\vert 10\right\rangle _{A}\left\vert 1\right\rangle
_{V}$, state after the computation $\rho_{2}\operatorname{e}^{i\varphi_{2}%
}\left\vert 0011\right\rangle _{B}\left\vert 10\right\rangle _{A}\left\vert
0\right\rangle _{V}$.\ If she performs function evaluation for $\mathbf{a}=11$
instead, this originates other 4 histories, etc. If we sum together all the
different histories, each with a suitable phase, and normalize, we obtain the
function evaluation stage of the quantum algorithm.

To obtain the quantum algorithm, we should choose history phases and the final
unitary transformation applied to register $A$ in such a way that the
information about the solution readable in that register at the end of the
algorithm is maximized.

\subsection{Simon's and the hidden subgroup algorithms}

In Simon's $\left[  8\right]  $ algorithm, the set of functions is all the
$f_{\mathbf{b}}:\left\{  0,1\right\}  ^{n}\rightarrow\left\{  0,1\right\}
^{n-1}$ such that $f_{\mathbf{b}}\left(  \mathbf{a}\right)  =f_{\mathbf{b}%
}\left(  \mathbf{c}\right)  $ if and only if $\mathbf{a}=\mathbf{c}$\ or
$\mathbf{a}=\mathbf{c}\oplus\mathbf{h}^{\left(  \mathbf{b}\right)  }$;
$\oplus$\ denotes bitwise modulo 2 addition; the bit string $\mathbf{h}%
^{\left(  \mathbf{b}\right)  }\mathbf{\equiv~}h_{0}^{\left(  \mathbf{b}%
\right)  },h_{1}^{\left(  \mathbf{b}\right)  },...,h_{n-1}^{\left(
\mathbf{b}\right)  }$, depending on $\mathbf{b}$ and belonging to $\left\{
0,1\right\}  ^{n}$ excluded the all zeroes string, is a sort of period of the
function. Table (\ref{periodic}) gives the set of functions for $n=2$. The bit
string $\mathbf{b}$ is both the suffix and the table of the function. Since
$\mathbf{h}^{\left(  \mathbf{b}\right)  }\oplus\mathbf{h}^{\left(
\mathbf{b}\right)  }=0$, each value of the function appears exactly twice in
the table, thus 50\% of the rows plus one surely identify $\mathbf{h}^{\left(
\mathbf{b}\right)  }$.
\begin{equation}%
\begin{tabular}
[c]{|c|c|c|c|c|c|c|}\hline
& $\mathbf{h}^{\left(  0011\right)  }=01$ & $\mathbf{h}^{\left(  1100\right)
}=01$ & $\mathbf{h}^{\left(  0101\right)  }=10$ & $\mathbf{h}^{\left(
1010\right)  }=10$ & $\mathbf{h}^{\left(  0110\right)  }=11$ & $\mathbf{h}%
^{\left(  1001\right)  }=11$\\\hline
$\mathbf{a}$ & $f_{0011}\left(  \mathbf{a}\right)  $ & $f_{1100}\left(
\mathbf{a}\right)  $ & $f_{0101}\left(  \mathbf{a}\right)  $ & $f_{1010}%
\left(  \mathbf{a}\right)  $ & $f_{0110}\left(  \mathbf{a}\right)  $ &
$f_{1001}\left(  \mathbf{a}\right)  $\\\hline
00 & 0 & 1 & 0 & 1 & 0 & 1\\\hline
01 & 0 & 1 & 1 & 0 & 1 & 0\\\hline
10 & 1 & 0 & 0 & 1 & 1 & 0\\\hline
11 & 1 & 0 & 1 & 0 & 0 & 1\\\hline
\end{tabular}
\ \label{periodic}%
\end{equation}

Bob selects a value of $\mathbf{b}$. Alice's problem is finding the value of
$\mathbf{h}^{\left(  \mathbf{b}\right)  }$, "hidden" in $f_{\mathbf{b}}\left(
\mathbf{a}\right)  $, by computing $f_{\mathbf{b}}\left(  \mathbf{a}\right)
=f\left(  \mathbf{b},\mathbf{a}\right)  $ for different values of $\mathbf{a}%
$. In present knowledge, a classical algorithm requires a number of
computations of $f\left(  \mathbf{b},\mathbf{a}\right)  $ exponential in $n$.
The quantum algorithm solves the hard part of this problem, namely finding a
string $\mathbf{s}_{j}^{\left(  \mathbf{b}\right)  }$ orthogonal\footnote{The
modulo 2 addition of the bits of the bitwise product of the two strings should
be zero.} to $\mathbf{h}^{\left(  \mathbf{b}\right)  }$, with one computation
of $f\left(  \mathbf{b},\mathbf{a}\right)  $. There are $2^{n-1}$ such
strings. Running the quantum algorithm yields one of these strings at random
(see further below). The quantum algorithm is iterated until finding $n-1$
different strings. This allows to find $\mathbf{h}^{\left(  \mathbf{b}\right)
}$ by solving a system of modulo 2 linear equations. Register $B$\ is now
$2^{n}\left(  n-1\right)  $-qubit, given that $\mathbf{b}$ is the sequence of
$2^{n}$ fields each on $n-1$\ bits.

The initial state of the algorithm relativized to Alice, with register $V$
prepared in the all zeroes string (just one zero for $n=2$), is:%

\begin{align}
&  \frac{1}{2}\left(  \rho_{0}\operatorname{e}^{i\varphi_{0}}\left\vert
0011\right\rangle _{B}+\rho_{1}\operatorname{e}^{i\varphi_{1}}\left\vert
1100\right\rangle _{B}+\rho_{2}\operatorname{e}^{i\varphi_{2}}\left\vert
0101\right\rangle _{B}+\rho_{3}\operatorname{e}^{i\varphi_{3}}\left\vert
1010\right\rangle _{B}+...\right) \label{insimon}\\
&  \left(  \left\vert 00\right\rangle _{A}+\left\vert 01\right\rangle
_{A}+\left\vert 10\right\rangle _{A}+\left\vert 11\right\rangle _{A}\right)
\left\vert 0\right\rangle _{V}.\nonumber
\end{align}
\ Computing $f\left(  \mathbf{b},\mathbf{a}\right)  $, which changes the
content of $V$ from zero to the outcome of the computation, and performing the
Hadamard transform on register $A$ yields:%

\begin{equation}
\frac{1}{2}\left\{
\begin{array}
[c]{c}%
(\rho_{0}\operatorname{e}^{i\varphi_{0}}\left\vert 0011\right\rangle _{B}%
+\rho_{1}\operatorname{e}^{i\varphi_{1}}\left\vert 1100\right\rangle
_{B})\left[  (\left\vert 00\right\rangle _{A}+\left\vert 10\right\rangle
_{A})\left\vert 0\right\rangle _{V}+(\left\vert 00\right\rangle _{A}%
-\left\vert 10\right\rangle _{A})\left\vert 1\right\rangle _{V}\right] \\
+(\rho_{2}\operatorname{e}^{i\varphi_{2}}\left\vert 0101\right\rangle
_{B}+\rho_{3}\operatorname{e}^{i\varphi_{3}}\left\vert 1010\right\rangle
_{B})\left[  (\left\vert 00\right\rangle _{A}+\left\vert 01\right\rangle
_{A})\left\vert 0\right\rangle _{V}+(\left\vert 00\right\rangle _{A}%
-\left\vert 01\right\rangle _{A})\left\vert 1\right\rangle _{V}\right]  +...
\end{array}
\right\}  , \label{hsimon}%
\end{equation}
where, for each value of $\mathbf{b}$, register $A$ (no matter the content of
$V$) hosts even weighted\ superpositions of the $2^{n-1}$ strings
$\mathbf{s}_{j}^{\left(  \mathbf{b}\right)  }$ orthogonal to $\mathbf{h}%
^{\left(  \mathbf{b}\right)  }$. By measuring $\hat{A}$\ and $\hat{B}$ in
state (\ref{hsimon}), we obtain at random Bob's selection of $\mathbf{b}$ and
one of the $\mathbf{s}_{j}^{\left(  \mathbf{b}\right)  }$.

We leave $B$ in its after-measurement state, thus fixing the value of
$\mathbf{b}$, and iterate the "right part" of the algorithm (preparation of
registers $A$\ and $V$, computation of $f\left(  \mathbf{b},\mathbf{a}\right)
$, and measurement of $\hat{A}$) until obtaining $n-1$ different
$\mathbf{s}_{j}^{\left(  \mathbf{b}\right)  }$.

We check that the quantum algorithm requires the number of function
evaluations of a classical algorithm that knows in advance 50\% of
$\mathcal{I}$. Any $\mathbf{s}_{j}^{\left(  \mathbf{b}\right)  }$ is a
solution of the problem addressed by the quantum part of Simon's algorithm.
The advanced information is any 50\% of the information about the solution
contained in $\mathbf{b}$. For reasons of symmetry, this is any 50\% of the
table of the function that does not contain the same value of the function
twice. In fact, the half tables that contain a same value twice already
specify the value of $\mathbf{h}^{\left(  \mathbf{b}\right)  }$ and thus the
value of any $\mathbf{s}_{j}^{\left(  \mathbf{b}\right)  }$. For the half
tables that do not contain the same value of the function twice, the solution
is always identified by computing $f\left(  \mathbf{b},\mathbf{a}\right)  $
for only one value of $\mathbf{a}$\ (any one) outside the half table. The new
value of the function is necessarily a value already present in the half
table, which identifies $\mathbf{h}^{\left(  \mathbf{b}\right)  }$ and thus
all the $\mathbf{s}_{j}^{\left(  \mathbf{b}\right)  }$. Thus, both the quantum
algorithm and the advanced information classical algorithm require just one
function evaluation.

As in section 3.1, the above defined advanced information is back evolved
projection on the solution and cannot exceed 50\% of $\mathcal{I}$.

Now we go to the history superposition picture.\ For example, let us assume
that the advanced information is $f\left(  \mathbf{b},00\right)  =0$ and
$f\left(  \mathbf{b},11\right)  =1$, namely the first and last row of either
$f_{0011}\left(  \mathbf{a}\right)  $ or $f_{0101}\left(  \mathbf{a}\right)  $
-- see table (\ref{periodic}). To find which is the case, Alice should perform
function evaluation for either $\mathbf{a}=01$ or $\mathbf{a}=10$. We assume
she does it for $\mathbf{a}=01$. If the result of the computation is $0$, this
means that $\mathbf{b}=0011$. This originates two classical computation
histories in quantum notation: \# 1: initial state $\rho_{0}\operatorname{e}%
^{i\varphi_{0}}\left\vert 0011\right\rangle _{B}\left\vert 01\right\rangle
_{A}\left\vert 0\right\rangle _{V}$, state after the computation $\rho
_{0}\operatorname{e}^{i\varphi_{0}}\left\vert 0011\right\rangle _{B}\left\vert
01\right\rangle _{A}\left\vert 0\right\rangle _{V}$; \# 2: initial state
$\rho_{0}\operatorname{e}^{i\varphi_{0}}\left\vert 0011\right\rangle
_{B}\left\vert 01\right\rangle _{A}\left\vert 1\right\rangle _{V}$, state
after the computation $\rho_{0}\operatorname{e}^{i\varphi_{0}}\left\vert
0011\right\rangle _{B}\left\vert 01\right\rangle _{A}\left\vert 1\right\rangle
_{V}$. If the result of the computation is $1$, this means that $\mathbf{b}%
=0101$. This originates two histories: \# 3: initial state $\rho
_{2}\operatorname{e}^{i\varphi_{2}}\left\vert 0101\right\rangle _{B}\left\vert
01\right\rangle _{A}\left\vert 0\right\rangle _{V}$, state after the
computation $\rho_{2}\operatorname{e}^{i\varphi_{2}}\left\vert
0101\right\rangle _{B}\left\vert 01\right\rangle _{A}\left\vert 1\right\rangle
_{V}$; \# 3: initial state $\rho_{2}\operatorname{e}^{i\varphi_{2}}\left\vert
0101\right\rangle _{B}\left\vert 01\right\rangle _{A}\left\vert 1\right\rangle
_{V}$, state after the computation $\rho_{2}\operatorname{e}^{i\varphi_{2}%
}\left\vert 0101\right\rangle _{B}\left\vert 01\right\rangle _{A}\left\vert
0\right\rangle _{V}$.\ If she performs function evaluation for $\mathbf{a}=10$
instead, this originates other 4 histories, etc. If we sum together all the
different histories, each with a suitable phase, and normalize, we obtain the
function evaluation stage of the quantum algorithm.

To obtain the quantum algorithm, we should choose history phases and the final
unitary transformation applied to register $A$ in such a way that the
information about the solution readable in that register at the end of the
algorithm is maximized.

The 50\% rule also applies to the generalized Simon's problem and to the
hidden subgroup problem. In fact the corresponding algorithms are essentially
the same as the algorithm that solves Simon's problem. In the hidden subgroup
problem, the set of functions $f_{\mathbf{b}}:G\rightarrow W$ map a group $G$
to some finite set $W$\ with the property that there exists some subgroup
$S\leq G$ such that for any $\mathbf{a},\mathbf{c}\in G$, $f_{\mathbf{b}%
}\left(  \mathbf{a}\right)  =f_{\mathbf{b}}\left(  \mathbf{c}\right)  $ if and
only if $\mathbf{a}+S=\mathbf{c}+S$. The problem is to find the hidden
subgroup $S$ by computing $f_{\mathbf{b}}\left(  \mathbf{a}\right)  $ for
various values of $\mathbf{a}$. Now, a large variety of problems solvable with
a quantum speed-up can be re-formulated in terms of the hidden subgroup
problem $\left[  9\right]  $. Among these we find: Deutsch's problem, finding
orders, finding the period of a function (thus the problem solved by the
quantum part of Shor's factorization algorithm), discrete logarithms in any
group, hidden linear functions, self shift equivalent polynomials, Abelian
stabilizer problem, graph automorphism problem.

\section{Applying the 50\% rule to the search of quantum speed ups}

In hindsight, the quantum algorithms examined are skillfully designed around
the 50\% rule. In unstructured data base search, the advanced knowledge of
50\% of the solution yields a quadratic speed-up, given that the number of
function evaluations goes from $\operatorname{O}\left(  2^{n}\right)  $ to
$\operatorname{O}\left(  2^{n/2}\right)  $. Thus, the possibility of a
quadratic speed-up is established by the 50\% rule, one does not need to know
Grover's algorithm. Similarly, in the structured algorithms that yield an
exponential speed-up, the problem is chosen in such a way that, if one knows
in advance 50\% of the rows of the table of the function, computing
$f_{\mathbf{b}}\left(  \mathbf{a}\right)  $ for a single value of $\mathbf{a}%
$\ outside the half table yields the solution. Thus, the possibility of an
exponential speed-up is established by the 50\% rule before knowing the
quantum algorithm.

One way of searching for new quantum speed ups is thus looking for problems
solvable with a single computation of $f_{\mathbf{b}}\left(  \mathbf{a}%
\right)  $\ once that 50\% of the rows of the table of the function are known.
We provide an example -- see also Ref. $\left[  4\right]  $. The set of
functions is the $4!$\ functions $f_{\mathbf{b}}:\left\{  0,1\right\}
^{2}\rightarrow\left\{  0,1\right\}  ^{2}$ such that the sequence of function
values is a permutation of the values of the argument -- see table (\ref{perm}).%

\begin{equation}%
\begin{tabular}
[c]{|c|c|c|c|c|}\hline
$\mathbf{a}$ & $f_{00011110}\left(  \mathbf{a}\right)  $ & $f_{00110110}%
\left(  \mathbf{a}\right)  $ & $f_{00011011}\left(  \mathbf{a}\right)  $ &
$...$\\\hline
00 & 00 & 00 & 00 & ...\\\hline
01 & 01 & 11 & 01 & ...\\\hline
10 & 11 & 01 & 10 & ...\\\hline
11 & 10 & 10 & 11 & ...\\\hline
\end{tabular}
\label{perm}%
\end{equation}
The string $\mathbf{b}$\ is both the suffix and the table of the function --
the sequence of function values for increasing values of the argument. We have
chosen this set because, if we know 50\% of the rows of one table, we can
identify the corresponding value of $\mathbf{b}$\ with a single computation of
$f_{\mathbf{b}}\left(  \mathbf{a}\right)  $. Without advanced information,
three computations of $f_{\mathbf{b}}\left(  \mathbf{a}\right)  $ are
required. Thus there is room for a speed-up. We build a quantum algorithm over
this possibility. Register $B$\ is $8$ qubits, registers $A$ is $2$ qubits,
and register $V$ is $2$\ qubits, denoted $V_{0}$ and $V_{1}$. The result of
the computation of $f_{\mathbf{b}}\left(  \mathbf{a}\right)  =f\left(
\mathbf{b},\mathbf{a}\right)  $ is bitwise modulo $2$ added to the former
content of $V$. The initial state is:%

\begin{align*}
&  \frac{1}{8\sqrt{6}}\left(  \operatorname{e}^{i\varphi_{0}}\left\vert
00011110\right\rangle _{B}+\operatorname{e}^{i\varphi_{1}}\left\vert
00110110\right\rangle _{B}+\operatorname{e}^{i\varphi_{2}}\left\vert
00011011\right\rangle _{B}...\right) \\
&  \left(  \left\vert 00\right\rangle _{A}+\left\vert 01\right\rangle
_{A}+\left\vert 10\right\rangle _{A}+\left\vert 11\right\rangle _{A}\right)
\left(  \left\vert 0\right\rangle _{V_{0}}-\left\vert 1\right\rangle _{V_{0}%
}\right)  \left(  \left\vert 0\right\rangle _{V_{1}}-\left\vert 1\right\rangle
_{V_{1}}\right)  .
\end{align*}
Computing $f\left(  \mathbf{b},\mathbf{a}\right)  $, then performing the
Hadamard transform on register $A$, yields%

\begin{align*}
&  \frac{1}{4\sqrt{6}}\left[  \left(  \operatorname{e}^{i\varphi_{0}%
}\left\vert 00011110\right\rangle _{B}+~...\right)  \left\vert 01\right\rangle
_{A}+\left(  \operatorname{e}^{i\varphi_{1}}\left\vert 00110110\right\rangle
_{B}+~...\right)  \left\vert 10\right\rangle _{A}+\left(  \operatorname{e}%
^{i\varphi_{2}}\left\vert 00011011\right\rangle _{B}+~...\right)  \left\vert
11\right\rangle _{A}\right] \\
&  \left(  \left\vert 0\right\rangle _{V_{0}}-\left\vert 1\right\rangle
_{V_{0}}\right)  \left(  \left\vert 0\right\rangle _{V_{1}}-\left\vert
1\right\rangle _{V_{1}}\right)  ,
\end{align*}
an entangled state where three orthogonal states of $B$ (each a superposition
of $8$ values of $\mathbf{b}$, corresponding to a partition of the set of $24$
functions) are correlated with, respectively, $\left\vert 01\right\rangle
_{A},~\left\vert 10\right\rangle _{A},~$and $\left\vert 11\right\rangle _{A}$.
Measuring $\hat{A}$ in the above state tells which of the three partitions the
function belongs to. In the case of a classical algorithm, identifying the
partition requires three computations of $f\left(  \mathbf{b},\mathbf{a}%
\right)  $, as readily checked. There is thus a quantum speed-up.

With the 50\% rule, one can figure out any number of these speed ups in terms
of number of function evaluations. Thus, this rule provides a playground for
studying the engineering of quantum algorithms.

\section{Conclusion}

Summarizing, moving from classical to quantum problem solving, the classical
problem-solution correlation becomes quantum. There is quantum correlation
between the selection of an eigenvalue of $\hat{B}$ on the part of Bob and
that of an eigenvalue of $\hat{A}$ -- the solution -- on the part of Alice.
The random selection of an eigenvalue of $\hat{B}$ is required to set register
$B$\ in a known eigenstate, then transformed into the desired eigenstate by
means of a permutation of the basis vectors of $B$. From the standpoint of
correlation, which is defined on repetitions of the same quantum experiment,
this permutation should be considered fixed: the fact that Bob can change it
to always obtain the desired value of $\mathbf{b}$ belongs to a different
film. Because of quantum correlation, all is like Alice contributed to
selecting 50\% of the information that specifies the problem. As the solution
is a function of the problem, this becomes Alice knowing in advance 50\% of
the information that specifies the solution.

The fact that the quantum speed up comes from comparing two classical
algorithms, with and without advanced information, has a practical interest.
It allows to characterize the problems solvable with a quantum speed up in an
entirely computer science framework, with no physics involved -- an important
simplification. It should also allow to identify new quantum speed ups, as
exemplified in section 4.

The fact that quantum algorithms are quicker because they know in advance 50\%
of the solution they will themselves produce in the future has an obvious
interest from the standpoint of the philosophy of quantum mechanics.

Future work should aim to check the 50\% rule for all quantum algorithms found
so far, to possibly demonstrate it in a more general way, for example for the
generic quantum computational network or quantum Turing machine, and to
explore the quantum speed ups achievable on the basis of the 50\% rule.

\subsection*{Acknowledgements}

The author thanks David Finkelstein for useful discussions, David Deutsch, Tom
Toffoli, and Lev Vaidman for useful comments.

\subsection*{Bibliography}

$\left[  1\right]  $\ D. Gross, S. T. Flammia, and J. Eisert, Phys. Rev. Lett.
\textbf{102} (19) (2009).

$\left[  2\right]  $\ G. Castagnoli and D. Finkelstein, Proc. Roy. Soc. Lond.
A \textbf{457}, 1799 . arXiv:quant-ph/0010081 v1 (2001).

$\left[  3\right]  $\ G. Castagnoli, Int. J. Theor. Phys. vol. 48 issue 8,
2412 (2009).

$\left[  4\right]  $ G. Castagnoli, Int. J. Theor. Phys.,vol. 48 issue 12,
3383 (2009).

$\left[  5\right]  $\ L. K. Grover, Proc. 28th Ann. ACM Symp. Theory of
Computing (1996).

$\left[  6\right]  $\ \ C. Rovelli, Int. J. Theor. Phys. \textbf{35}, 1637 (1996).

$\left[  7\right]  $\ D. Deutsch and R. Jozsa, Proc. Roy. Soc. (Lond.) A,
\textbf{439}, 553 (1992).

$\left[  8\right]  $ D. Simon, Proc. 35th Ann. Symp. on Foundations of Comp.
Sci.,\textit{\ }116 (1994).

$\left[  9\right]  $ P. Kaye, R. Laflamme, and M. Mosca, \textit{An
introduction to Quantum Computing}, Oxford University Press, 146 (2007).

\end{document}